# Charge radii of atomic nuclei and shell effects

**M. V. Chushnyakova (1), I. I. Gontchar (2), E. V. Kulik (1)**

(1) Physics Department, Omsk State Technical University, Omsk, Russia
(2) Physics and Chemistry Department, Omsk State Transport University, Omsk, Russia

*e-mail: vigichar@hotmail.com*

In the present work, we study the experimental data on the root mean square charge radii of the spherical atomic nuclei. The purpose of this analysis is to clarify what is the manifestation of the neutron shells in the considered observable and to find regularities in the behavior of the charge radii when a shell neutron is crossed. As the result, we have found manifestations of the shells $N\_sh = 8, 28, 50$ and $82$. It turns out that the well-known shells with the neutron numbers 20 and 126 do not show up. The present analysis is performed by means of the relative charge radius $x(N, N\_sh, Z)$. This value is equal to the charge radius of a given isotope over the charge radius of the isotope of the given element with the closed neutron shell. It is shown that the values of $x$ form groups corresponding to different values of $N\_sh$. Prediction for unknown charge radius of sulfur-30 has been made.

Acknowledgments: The work was supported by the Foundation for the Advancement of Theoretical Physics and Mathematics "BASIS".

## 1. INTRODUCTION

The effects related to the proton- and neutron single particle shell structure of the energy levels in atomic nuclei are discussed in the literature since 1930s [1-3]. This structure manifests itself in the binding energies and shapes of the nuclei [4]. It is conclusively established that at the charge number $Z = 2, 8, 20, 28, 50, 82$ (so-called "magic numbers") the closed proton shells appear resulting in the maximum binding energy per nucleon by absolute value [5]. The neutron closed shells appear at the same magic neutron numbers $N$ and in addition at $N = 126$.

The single particle shell structure is expected to show up in other nuclear properties, e.g., to influence the shapes of nuclei. Indeed, it is firmly established experimentally that the nuclei having both $Z$ and $N$ magic possess spherical shape, e.g., $^{16}O, ^{40}Ca, ^{48}Ca, ^{208}Pb$ [5]. This is mostly confirmed theoretically [6], although there are certain contradictions. For instance, theory [6] predicts for $^{16}O$ non-zero values of the parameters of quadrupole $\beta_2$, octupole $\beta_3$, and hexadecapole $\beta_4$ deformations: $\beta_2 = -0{,}010$, $\beta_3 = -0{,}258$, $\beta_4 = -0{,}122$. Thus, one should be careful with theory [6] concerning the shape of nuclei.

The situation with the shapes of nuclei with $Z > 100$ ("superheavy nuclei") is even less definite: the corresponding experimental data are scares whereas theory [6] predicts almost spherical shape for nuclei with $114 < Z < 118$. There are mostly two kinds of experimental data from which the information of the nuclear shape is extracted. First, these are measured nuclear electric quadrupole moments [3,7] which in turn are obtained from the fine structure of atomic spectra [5]. Second, this is the structure of low-lying excited energy levels of nuclei. For even-even deformed nuclei, the rotational levels 2+, 4+, 6+ appear whose energies approximately are related as 1:3:6 whereas the rotational degree of freedom absents for spherical nuclei [5,8]. It is interesting to note that for neighboring nuclei with different symmetry types the first excited state 2+ is significantly higher for spherical nucleus than for the deformed one [9].

We did not manage finding a systematics of experimental data of nuclear deformations at the ground state which could be analyzed with respect to the regularities related to the shell effects. Yet there is perfect experimental review [10] where the rms nuclear charge radii are systematized. This is the goal of the present work to analyze quantitatively these data searching for regularities related to the shells.

## 2. ANALYSIS OF THE EXPERIMENTAL CHARGE EADII OF EVEN-EVEN NUCLEI

A quantitative analysis, aiming to separate the shell effects in any nuclear data, is challenging to certain extent. First, almost all nuclear characteristics are subjects of fluctuations related to odd-even effects. In the present work, for excluding these fluctuations, following the line of [11,12], we consider only even-even nuclei.

Second, in the first approximation, many nuclear characteristics (e.g., mass or volume) grow monotonically along the beta-stability line with the charge number, whereas the shell effects just slightly modulate this growth. One example illustrating this statement is shown in Fig. 1. The solid line for rms charge radius in Fig. 1 is built according to a formula from Ref. [10].

$$R_0 = \left(r_0 + \frac{r_1}{A^{2/3}} + \frac{r_2}{A^{4/3}}\right) \times A^{1/3}, \qquad (1)$$

where $A$ is the mass number, $r_0 = 0{,}907$ fm, $r_1 = 1{,}105$ fm, $r_2 = -0{,}548$ fm. The experimental values from [10] are shown by symbols.

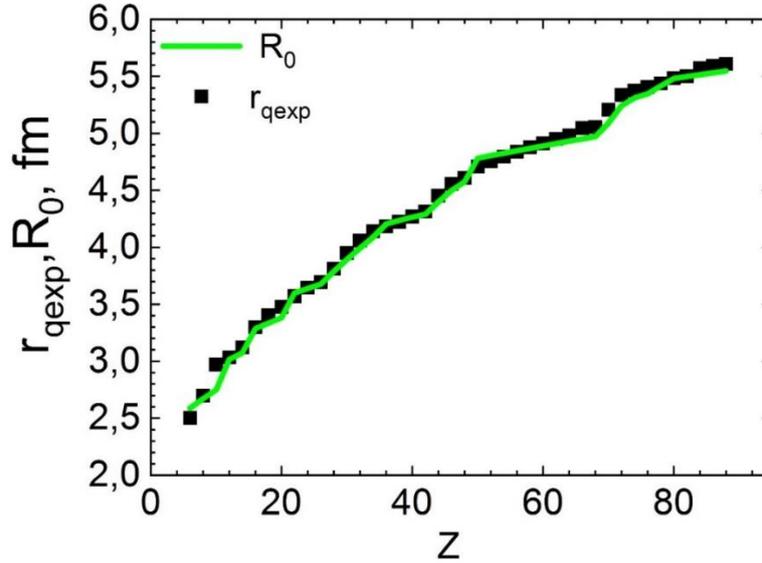

Fig. 1. Experimental (symbols) and analytical (see Eq. (1)) rms charge radii versus the charge number.

Table 1. Isotopes used for the analysis, their experimental rms charge radii with its error from Ref. [10], the halflifes from Ref. [9].

| El. | $A$ | $N$ | $r_{qexp}$, fm | $\Delta r_{qexp}$, fm | Halflife | El. | $A$ | $N$ | $r_{qexp}$, fm | $\Delta r_{qexp}$, fm | Halflife |
|---|---|---|---|---|---|---|---|---|---|---|---|
| $_6$C | 14 | 8 | 2,5025 | 0,0087 | $5{,}7 \cdot 10^3$ y | $_{54}$Xe | 136 | 82 | 4,7964 | 0,0047 | $> 3{,}60\ 10^{20}$ y |
| $_8$O | 16 | 8 | 2,6991 | 0,0052 | Stable | $_{56}$Ba | 138 | 82 | 4,8378 | 0,0046 | Stable |
| $_{10}$Ne | 18 | 8 | 2,9714 | 0,0076 | 1,67 s | $_{58}$Ce | 140 | 82 | 4,8771 | 0,0018 | Stable |
| $_{16}$S | 36 | 20 | 3,2985 | 0,0024 | Stable | $_{60}$Nd | 142 | 82 | 4,9123 | 0,0025 | Stable |
| $_{18}$Ar | 38 | 20 | 3,4028 | 0,0019 | Stable | $_{62}$Sm | 144 | 82 | 4,9524 | 0,0034 | Stable |
| $_{20}$Ca | 40 | 20 | 3,4776 | 0,0019 | Stable | $_{64}$Gd | 146 | 82 | 4,9801 | 0,0140 | 48,27 d |
| $_{22}$Ti | 50 | 28 | 3,5704 | 0,0022 | Stable | $_{64}$Dy | 148 | 82 | 5,0455 | 0,2389 | 3,30 m |
| $_{24}$Cr | 52 | 28 | 3,6452 | 0,0042 | Stable | $_{68}$Er | 150 | 82 | 5,0548 | 0,0254 | 18,50 s |
| $_{26}$Fe | 54 | 28 | 3,6933 | 0,0019 | Stable | $_{70}$Yb | 152 | 82 | 5,0423 | 0,0146 | 3,03 s |
| $_{36}$Kr | 86 | 50 | 4,1835 | 0,0021 | Stable | $_{80}$Hg | 206 | 126 | 5,4837 | 0,0040 | 8,32 m |
| $_{38}$Sr | 88 | 50 | 4,2240 | 0,0018 | Stable | $_{82}$Pb | 208 | 126 | 5,5012 | 0,0013 | Stable |
| $_{40}$Zr | 90 | 50 | 4,2694 | 0,0010 | Stable | $_{84}$Po | 210 | 126 | 5,5704 | 0,0176 | 138,38 d |
| $_{42}$Mo | 92 | 52 | 4,3151 | 0,0012 | Stable | $_{86}$Rn | 212 | 126 | 5,5915 | 0,0176 | 23,90 m |
| $_{50}$Sn | 132 | 82 | 4,7093 | 0,0076 | 39,7 s | $_{88}$Ra | 214 | 126 | 5,6079 | 0,0177 | 2,46 s |
| $_{52}$Te | 134 | 82 | 4,7569 | 0,0041 | 41,80 m | | | | | | |

To make the shell effects more visible one should exclude from consideration the monotonic component of the studied nuclear characteristic. For this aim we propose the following algorithm. Of all nuclei with even Z we have selected those for which: i) at least one isotope with closed neutron shell is known; ii) the experimental rms charge radius of this isotope is known. Corresponding information is comprised in Table 1 where the values of $r_{qexp}$ and their errors are taken from review [10]. For radioactive nuclei the halflifes are presented, too. Let us note that information of the electric quadrupole moments of these nuclei absents in review [7] making us to conclude that these nuclei are spherical.

In Fig. 2 we present the ratio
$$x = r_{qexp}/r_{qexpsh} \quad (2)$$
of the charge radius of an isotope of an element from Table 1 to the charge radius of the isotope with closed neutron shell $N = N_{sh}$ of this element. The values of $x$ are displayed in Fig. 2 versus $N - N_{sh}$ (see corresponding $N_{sh}$ in the panels).

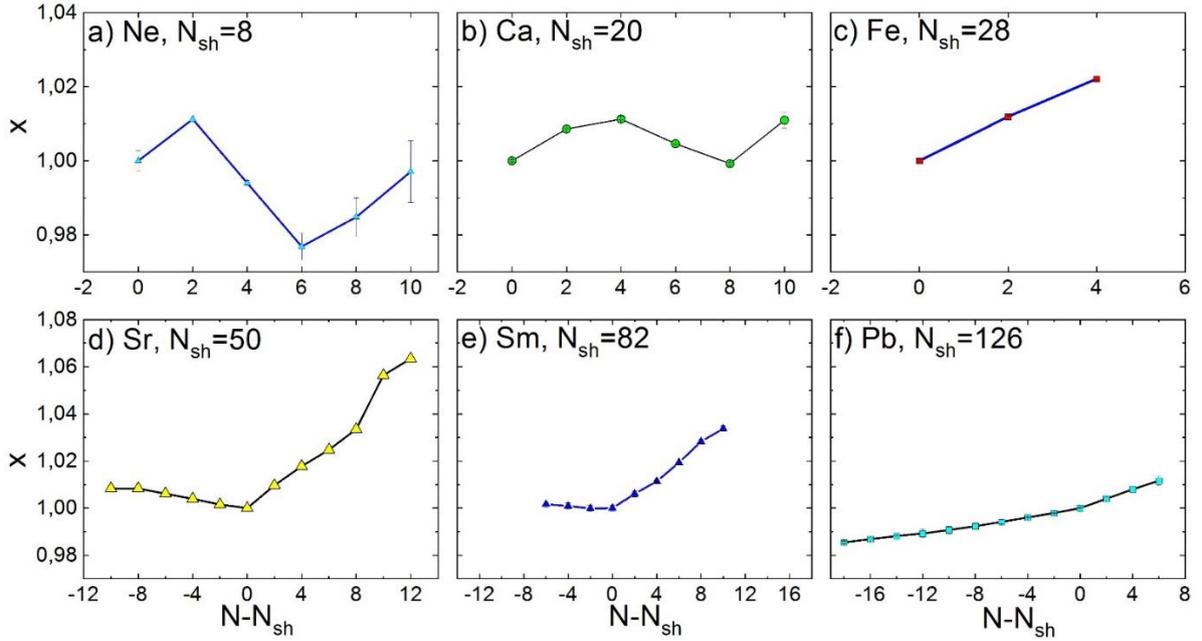

Fig. 2. Fractional charge radius $x$ (see Eq. (2)) versus $N - N_{sh}$ for Ne, Ca, Fe, Sr, Sm, Pb.

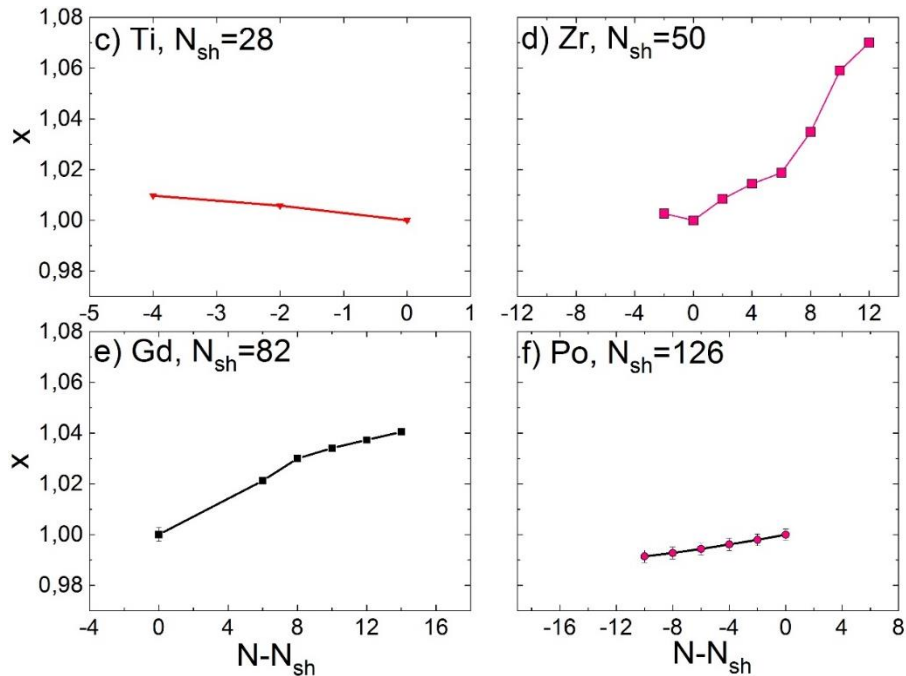

Fig. 3. Fractional charge radius $x$ (see Eq. (2)) versus $N - N_{sh}$ for Ti, Zr, Gd, Po.

We have failed finding any regularities in Fig. 2. Indeed, it is tempting to conclude that a closed neutron shell results to minimal charge radius (see Figs. $2d$), $2e$). It looks like Fig. $2b$) does not contradict to this statement. However, Figs. $2a$), $2f$) are in obvious contradiction to this conclusion.

In Fig. 3 we present the same dependences as in Fig. 2 but for the elements with Z larger by 2 units than in Fig. 2. There are two less panels because for the shells with $N_{sh} = 8, 20$ there are not enough number of even isotopes (less than 3). For convenience we keep the characters for the panels of Figs. 2 and 3 unchanged, i.e. one should compare the panels with the same characters. The curves in Figs. $2c$), $3c$) and Figs. $2e$), $3e$) do not contradict to each other although they do not look similar. They produce an impression of the pieces of the same broken line which supplement each other. The curves in Figs. $2d$), $3d$) and $2f$), $3f$) by pairs look rather similar.

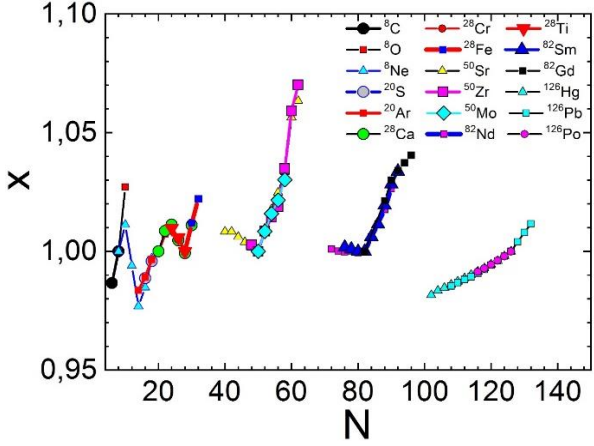

Fig. 4. $x$ versus N for all elements of Table 1.

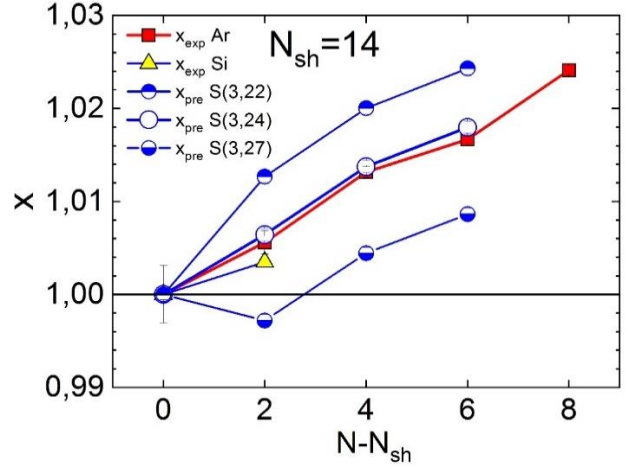

Fig. 5. $x$ versus $N - N_{sh}$: the experimental values for argon and silicon, the predicted values for sulfur. The latter are indicated in the brackets in units of fm. See text for details.

Let us now collect in Fig. 4 the values of $x$ versus the total neutron number for all elements of Table 1. The data corresponding to different neutron shells become grouped resembling parts of a single curve. This hypothetical curve resembles the well-known dependence of the atom ionization energy upon the atomic number [13] which is used in the textbooks for illustration of the Periodic law. It is interesting that well expressed minima correspond to the closed neutron shells $N_{sh} = 28, 50, 82$ whereas for the shells $N_{sh} = 20, 126$ no minima are observed. Yet the "quasi-shell" $N = 14$ indicated in review [10] is clearly seen.

As the next step, basing on the above analysis, we undertake an effort to predict the rms charge radius of an isotope for which the experimental data absent. Results of this effort are presented in Fig. 5 whose design is the same as of Figs. 2 and 3. To facilitate understanding of Fig. 5, the data used for this figure are comprised in Table 2. To make the experimental values of $x$ known for argon and silicon we accept the quasi-shell $N = 14$.

Table 2. Isotopes used in Fig. 5, their neutron numbers N, experimental nuclear rms charge radii $r_{qexp}$ with their errors $\Delta r_{qexp}$ from [10], the fractional charge radii $x$ (see Eq. (2)).

| Isotope | N | $r_{qexp}$, фм | $\Delta r_{qexp}$, фм | $x$ |
|---|---|---|---|---|
| $^{28}$Si | 14 | 3,1223 | 0,0024 | 1 |
| $^{30}$Si | 16 | 3,1332 | 0,0040 | 1,00349 |
| $^{32}$Ar | 14 | 3,3462 | 0,0103 | 1 |
| $^{34}$Ar | 16 | 3,3649 | 0,0043 | 1,00559 |
| $^{36}$Ar | 18 | 3,3902 | 0,002 | 1,01315 |
| $^{38}$Ar | 20 | 3,402 | 0,0017 | 1,01668 |
| $^{40}$Ar | 22 | 3,4269 | 0,0017 | 1,02412 |
| $^{30}$S | 14 | -- | -- | -- |
| $^{32}$S | 16 | 3,2608 | 0,0021 | 1,00642 |
| $^{34}$S | 18 | 3,2845 | 0,0021 | 1,01373 |
| $^{36}$S | 20 | 3,2982 | 0,0021 | 1,01796 |

The experimental rms charge radius of sulfur-30 is unknown although the isotope itself is known, its halflife is 1,18 s [9]. For some isotopes with similar halflifes the charge radii have been measured: $^{18}$Ne with the halflife 1,67 s represents just one example [9,10]. According to the revealed regularity, the curves $x(N-14)$ for silicon, sulfur, and argon should coincide. Therefore, we vary the presumable charge radius $r_{qpresh}$ («predicted shell») of $^{30}$S striving for such coincidence. In Fig. 5 one sees that the values of $x_{pre} = r_{qexp}/r_{qpresh}$ at $r_{qpresh} = 3,27$ fm and 3,22 фм deviate significantly from the common curve $x_{exp} = r_{qexp}/r_{qexpsh}$ for argon and silicon. Yet the value $r_{qpresh} = 3,24$ fm produces the sought coincidence. Thus, basing on the found regularity, we predict the experimental charge radius $r_{qpresh} = 3,240 \pm 0,015$ fm for $^{30}$S.

## 3. CONCLUSIONS

In the present work we have undertaken an effort to find regularities in nuclear rms charge radii near the neutron shells. For this analysis we have constructed a fractional charge radius $x$ (see Eq. (2)). It has turned out that in such representation the shells $N_{sh} = 8, 28, 50, 82$ are clearly observed whereas the neutron shells with neutron numbers 20 and 126 are not seen. Yet the quasi-shell N=14 discussed earlier in the literature is well expressed. The principal result of our work is that the values of $x$ form groups corresponding to different values of $N_{sh}$. The $x(N)$-dependencies corresponding to different Z almost indistinguishable within one group. In general, the quasi-curve $x(N)$ resembles the well-known dependence of the ionization energy upon the atomic number of a chemical element. Starting our work, we expected something like this because in both cases one deals with the manifestation of the shell structure of a single particle spectrum. Basing on the revealed regularities we have predicted the nuclear rms charge radius of the isotope $^{30}$S for which the experimental value is unknown.